# SPROBES: Enforcing Kernel Code Integrity on the TrustZone Architecture

Xinyang Ge, Hayawardh Vijayakumar, and Trent Jaeger
System and Internet Infrastructure Security Laboratory
The Pennsylvania State University
{xxg113, hvijay, tjaeger}@cse.psu.edu

*Abstract*—Many smartphones now deploy conventional operating systems, so the rootkit attacks so prevalent on desktop and server systems are now a threat to smartphones. While researchers have advocated using virtualization to detect and prevent attacks on operating systems (e.g., VM introspection and trusted virtual domains), virtualization is not practical on smartphone systems due to the lack of virtualization support and/or the expense of virtualization. Current smartphone processors do have hardware support for running a protected environment, such as the ARM TrustZone extensions, but such hardware does not control the operating system operations sufficiently to enable VM introspection. In particular, a conventional operating system running with TrustZone still retains full control of memory management, which a rootkit can use to prevent traps on sensitive instructions or memory accesses necessary for effective introspection. In this paper, we present SPROBES, a novel primitive that enables introspection of operating systems running on ARM TrustZone hardware. Using SPROBES, an introspection mechanism protected by TrustZone can instrument individual operating system instructions of its choice, receiving an unforgeable trap whenever any SPROBE is executed. The key challenge in designing SPROBES is preventing the rootkit from removing them, but we identify a set of five invariants whose enforcement is sufficient to restrict rootkits to execute only approved, SPROBE-injected kernel code. We implemented a proof-of-concept version of SPROBES for the ARM Fast Models emulator, demonstrating that in Linux kernel 2.6.38, only 12 SPROBES are sufficient to enforce all five of these invariants. With SPROBES we show that it is possible to leverage the limited TrustZone extensions to limit conventional kernel execution to approved code comprehensively.

## I. INTRODUCTION

Kernel rootkits pose a serious security threat because an adversary that successfully installs a rootkit can control all user-space processes running on that kernel and can hide from traditional antivirus software. Rootkits are a common problem for conventional operating systems, which are large, complex software systems that may contain many latent vulnerabilities and which run many processes with full privilege that if compromised can easily install a rookit.

Many smartphones have now adopted conventional operating systems to leverage existing functionality and hardware support. However, this makes rootkits a threat to smartphone systems as well. For example, CVE-2011-1823 [1] was assigned to an integer-overflow bug in a daemon process on Android 3.0. This bug enables a local adversary to gain root privilege, which is sufficient to install a rootkit. Thus, it is likely that rootkits for smartphones will be seen in the wild. The problem we examine in this paper is whether we can leverage available smartphone hardware to restrict and possibly detect rootkits already injected into a running kernel.

To restrict and/or detect kernel rootkits, researchers have proposed utilizing hardware and software mechanisms, such as virtualization and secure coprocessors, that have control over the operating system itself. Using such mechanisms, privileged operations, i.e., those can affect the actual processor state, are trapped providing the capability (e.g., by inspecting trap events) to monitor the operating system activities. For example, Garfinkel *et al.* [2] and Chen *et al.* [3] proposed using virtualization to monitor running operating systems. A hypervisor or VMM has visibility into a running operating system and can control operating system access to hardware resources, such as memory, enabling intrusion detection software to detect possible attacks, restricting the adversary to attacks that avoid detection. Many virtualization-based methods have been proposed to detect possible rootkit behaviors [4], [5], [6], [7], [8], [9]. In addition, researchers have also proposed methods to protect the integrity of kernel using VM introspection. SecVisor [10] and NICKLE [11] are two representative VMM-based solutions that claim to ensure integrity of kernel code over the system lifetime. However, virtualization is currently considered to be too expensive for smartphone systems. Also, coprocessor-based intrusion detection methods have been proposed [12], [13], where a separate hardware device with access to host memory can examine that memory with the aim of detecting rootkit behaviors. A limitation of coprocessor-based solutions is the coprocessor has a limited view of kernel state, in particular it cannot access registers of the host processor. Even the Intel System Management Mode (SMM) layer has been utilized to explore rootkit detection methods [14]. However, SMM environment is very resource limited, making introspection complex, and it is only available on Intel processors.

Based on the observation that over 95% smartphones now are using ARM processors [15], we envision that smartphones need an introspection solution for that hardware. Instead of virtualization extensions, ARM introduced hardware extensions for security called ARM TrustZone technology [16], [17] in ARMv6. Intuitively, TrustZone physically partitions all system resources (e.g., physical memory, peripherals, etc.) into two worlds: a *secure world* for security-sensitive resources and a *normal world* for conventional processing. TrustZone protects the secure world resources from the normal world, but the secure world can access resources in the normal world. This hardware separation protects the confidentiality and integrity of any computation in the secure world while permitting the secure world to view the normal world. One well-known use case of TrustZone is Apple's Touch ID. Using TrustZone, Apple established a trusted path between a fingerprint scanner and the secure world, which ensures that fingerprint database is protected from the rest of software [18]. Although the secure world can be used as a slave easily, making it a master can be troublesome because each world has full discretion over its own resources, meaning the normal world can use its resource (e.g., modify virtual memory settings) without mediation by the secure world.

In this paper, we utilize the TrustZone extensions to develop SPROBES, a novel instrumentation mechanism that enables the secure world to cause the normal world to trap on any normal world instruction and provide an unforgeable view of the normal world's processor state. This property of SPROBES helps facilitate monitoring over the normal world, as the secure world can choose the normal world instructions for which it wants to be notified, receive the current processor state of the normal world, and perform its desired monitoring actions before returning control to the normal world. Other than the placement of instrumentation, SPROBES are invisible to the normal world, so no changes are required to the operating system to utilize SPROBES.

We demonstrate SPROBES by developing a methodology for restricting the normal world's kernel execution to approved kernel code memory. To do this, we define a set of five invariants that when enforced imply that the supervisor mode of the normal world complies with the W$\oplus$X invariant [19] for an approved set of immutable kernel code pages, even if a rootkit has control of the normal world kernel. That is, adversaries running a rootkit in the normal world cannot inject new code without detection nor modify kernel code pages. As a by-product, SPROBES protect themselves from modification, even from a live rootkit. We show that these invariants can be enforced comprehensively using only 12 SPROBES for the Linux 2.6.38 kernel. We find that each SPROBE hit causes 5611 instructions to be executed using the ARM Fast Models emulator, but that most SPROBES are never hit in normal execution and those that are hit are either those enforced in normal VM introspection or account for less than 2% of the instructions executed.

**Contributions.** In this paper, we develop a TrustZone-based solution to restrict the executable memory available to live rootkits. In particular, we make the following contributions:

- We present SPROBES, a novel, cross-world instrumentation mechanism that can break on any normal world instruction transparently. The SPROBES mechanism enables the secure world to dynamically break into any normal world kernel routine and specify a trusted handler in the secure world to mediate that routine.

- We show that SPROBES can be used to restrict normal world kernel execution to only approved kernel code pages, even if that kernel is under the control of a rootkit. We identify five invariants that must be enforced and describe a placement strategy for SPROBES to enforce those invariants for ARM TrustZone architectures.

- We evaluate SPROBES by applying the placement strategy to the Linux 2.6.38 kernel running in the normal world. We find that only 12 SPROBES are necessary to enforce the five invariants comprehensively. Further, we find that such monitoring can be efficient, as most SPROBES are not hit in normal operation, and others account for less than 2% of the instructions executed or are typically applied by VM introspection.

With SPROBES we show that it is possible to leverage the limited ARM TrustZone extensions to limit conventional kernel execution to approved code comprehensively. Effectively, SPROBES enable implementation of the breakpoints necessary to perform typical VM introspection without a separate hypervisor layer.

The remainder of this paper is organized as follows. Section II introduces the problem of restricting kernel execution to approved code pages even when that kernel may be infected with a rootkit in terms of five invariants. Section III introduces the TrustZone architecture and explains the challenge of mediating normal world execution from the secure world. Section IV describes the design of SPROBES, a mechanism for setting breakpoints for monitoring normal world operating system execution. We develop a design that utilizes SPROBES to enforce the five invariants necessary to restrict kernel execution to approved code pages in Section V. We describe our implementation for the ARM Cortex-A15 processor emulated by Fast Models 8.1 emulator in Section VI. We evaluate the security guarantees achieved and performance overheads in Section VII. Section VIII presents related work. Finally, we conclude the paper in Section IX.

## II. PROBLEM OVERVIEW

Smartphone systems are now deployed on conventional operating systems, such as Linux or Windows, inheriting both their functionality and their threats. One significant threat is that an adversary may be capable of installing a kernel rootkit, giving the adversary full control of the operating system's execution. Conventional operating systems have large code bases, so latent, exploitable vulnerabilities are likely. Moreover, these systems have many privileged processes that all can install a rootkit trivially if compromised. As mentioned in the Introduction, vulnerabilities in privileged processes in Android systems have been reported.

A goal of system defenses is to restrict the attack options available to an adversary. Operating systems now deploy several defenses, such as W$\oplus$X and address space layout randomization [20], [19], [21], to prevent adversaries from using injected code as part of an attack and to increase the difficulty of guessing the location of existing code pages. W$\oplus$X limits adversaries to choose to either use a memory page as data, which can be written, or as code, which can be executed. If the defender selects only legitimate code pages as executable, then the adversary can only execute memory on those pages, limiting the adversaries' code available for running exploits. However, even with this limitation, an adversary can still launch attacks that reuse existing code, based on the idea of *return-oriented programming* [22] (ROP). ROP attacks leverage control of the stack pointer to execute exploits utilizing available code (i.e., set as executable) to implement the malicious logic. Address space layout randomization aims to make it impractical for an adversary to guess the correct address of available code, but an adversary may have means to extract the correct addresses, such as through information disclosure attacks [23] to enable launching of effective ROP-style attacks.

For this problem, we assume that the kernel initially is enforcing W⊕X over an approved set of kernel code pages[1], but that an adversary may still be capable of executing some form of a ROP attack to launch their rootkit. We observe that rootkits could compromise the integrity of kernel execution in the following ways. First, as W⊕X is the common technique used by the kernel to prevent code injection attacks, a rootkit can simply disable the W⊕X protection. For ARM processors, such as those used by most smartphones, this is done by disabling the *Write eXecute-Never* (WXN) bit. When the WXN bit is set, writable pages are never executable, regardless of how the page table is configured. In this case, we assume that the kernel has the WXN bit set initially, but a rootkit may execute existing code, if available, to disable that protection, enabling the rootkit to inject code in the kernel.

Second, to bypass the W⊕X protection, rootkits can modify page table entries. Suppose the adversary wants to modify a code page, which is initially read-only. Firstly, she alters the permission bits of that page from executable to writable, as they are mutually exclusive. Then, she may write to that page arbitrarily. Then, she changes the permission of that page from writable to executable, as if nothing has ever happened.

Third, an alternative approach from modifying page table entries in place is to duplicate a page table elsewhere and reset the page table base (e.g., TTBR on ARM and CR3 on x86). Following similar steps of the second approach, i.e., mark a page as writable and revert it to executable once the write is complete, the adversary would be able to inject kernel code.

Fourth, if the adversary can disable the MMU, she can bypass all the existing memory protections (e.g., page permission and W⊕X) as all of them are based on virtual memory system. Note that disabling MMU can limit what a rootkit can do as well. Disabling the MMU reduces the adversary to utilizing physical memory addresses, and she may not know the physical memory addresses of code necessary to continue her ROP attack. However, if the operating system maps the virtual addresses of kernel space to the exactly same physical addresses, the rootkit could benefit from disabling MMU without limiting its capabilities.

Lastly, an adversary may direct the kernel to execute instructions in user space instead. Since an adversary often controls user space (e.g., a root process) prior kernel exploitation, she can simply prepare the malicious instructions there and invoke them from kernel. This is possible because most operating systems (e.g., Linux) map kernel space and user space into one unified virtual address space, and page permissions are set in such a way that the kernel space has a one-way view of the user space.

To summarize, we cast the problem into the following security requirements:

- **S1**: Execution of user space code from the kernel must never be allowed.
- **S2**: W⊕X protection employed by the operating system must always be enabled.
- **S3**: The page table base address must always correspond to a legitimate page table.
- **S4**: Any modification to the page table entry must be mediated and verified.
- **S5**: MMU must be kept enabled to ensure all existing memory protections function properly.

### III. BACKGROUND: TRUSTZONE ARCHITECTURE

TrustZone [16] is a set of security extensions first added to ARMv6 processors. Its goal is to provide a secure, separate environment that protects the confidentiality and integrity of critical computation from conventional computation. TrustZone partitions both hardware and software resources into two worlds - the *secure world* for assets that are security-sensitive and the *normal world* for conventional processing.

The TrustZone hardware architecture is illustrated as Fig. 1. The processor core implements the two separate worlds, i.e., the normal world and the secure world. And it can be in one world at a time, meaning the two worlds are running in a time-sliced fashion. To maintain the processor state during the world switch, TrustZone adds a monitor mode, which resides only in the secure world. The software in the monitor mode ensures the state of the world (e.g., registers) that processor is leaving is saved, and the state of the world that processor is switching to is correctly restored. This procedure is similar to context switch between processes except there are some banked registers that are not required to be saved. The mechanisms by which the processor can enter monitor mode are tightly controlled. Interrupts can be configured to be handled in either the normal world or the secure world. In addition, the normal world may proactively execute a *Secure Monitor Call* (SMC), which is a dedicated instruction that can trigger the entry to monitor mode (i.e., the secure world). For ease of understanding, the SMC instruction is similar to "*int 0x80*" on Intel x86 and "*svc*" on ARM in terms of privilege mode switch.

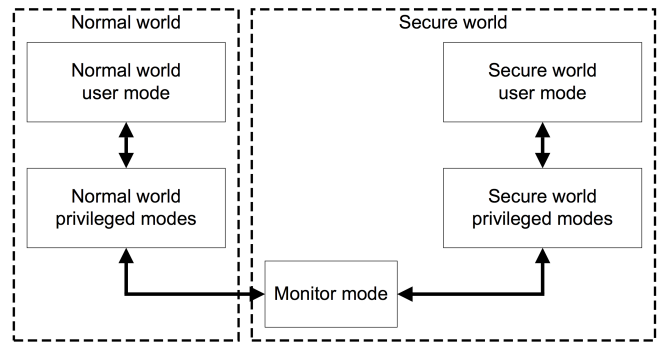

Fig. 1: TrustZone hardware architecture (from Figure 3.1 in [16])

The current world in which the processor runs is determined by the *Non-Secure* (NS) bit. In addition, almost all the system resources (e.g., memory, peripherals) are tagged with their own NS bits, determining the world they belong to. A general access control policy enforced by TrustZone is that the processor is able to access all the resources when running in the secure world, while it can only access normal world resources (i.e., those with the NS bit set) when running in the normal world. For example, memory hardware is partitioned into the two worlds. When the processor is in the normal world, it can only see the physical memory of its own world. After entering the secure world, the processor can see all the physical memory in the system.

---
[1]Note that even code pages may have read-only data embedded in them.

Unfortunately, the secure world as provided by the TrustZone architecture does not help much on protecting the kernel running in the normal world from rootkits. Unlike a VMM, the secure world is not more privileged than the normal world regarding the normal world resources. Once a hardware resource is assigned to the normal world, the secure world cannot control its access (e.g., by managing all physical memory, as a VMM would). For instance, the normal world has full privilege over its memory system, meaning it can arbitrarily set its virtual memory environment (i.e., virtual address mappings and page permissions) and access its physical memory without requiring any permission from the secure world. Another example is that interrupts assigned to the normal world are handled locally, without any secure world code being executed. By relinquishing this control to the normal world, rootkits can then tamper with the normal world's virtual memory environment as stated in Section II without being detected by the secure world.

## IV. SPROBES MECHANISM

To enable the secure world to control the execution of normal world events of its choice, we present SPROBES, an instrumentation mechanism that can transparently break on any instruction in the normal world. The control flow of an SPROBE is shown in Fig. 2. When an SPROBE is hit, the secure world immediately takes over the control, switches the context and invokes the specified SPROBE handler. The processor state of the normal world (e.g., register values when SPROBE is hit) is packed as a structure and passed to the invoked handler. Since this parameter cannot be forged by the normal world, the handler obtains a true view of the normal world from it. Finally, the control returns to the location where SPROBE is hit.

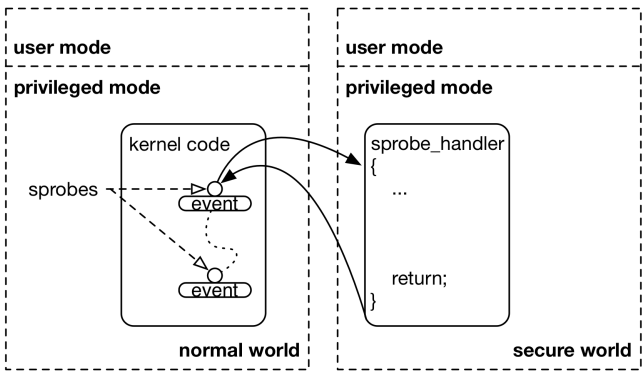

Fig. 2: SPROBE control flow

To trigger the secure world from a normal world instruction, we rewrite the specified instruction in the normal world as illustrated in Fig. 3. In this example, a secure world process (called a "trustlet") inserts an SPROBE at the normal world instruction of "*mov pc, lr*" (e.g., return from a function) by rewriting it to be the SMC instruction. The SMC instruction is the only instruction for triggering entry to the secure world as stated in Section III. Although code pages in the normal world might be write-protected, the secure world may rewrite any normal world instruction without causing an exception because the secure world has a different translation regime. This means all the virtual memory protections (e.g., W$\oplus$X) employed in the normal world are not applicable to the secure world. The secure world is authorized to access all physical memory without intervention of the normal world virtual memory system. Later, when the SPROBE is hit, meaning the program counter reaches the SMC instruction, the processor state will be switched to the secure world immediately. Within the secure world, the SPROBE handler is invoked, which can perform monitoring operations such as checking the state of the normal world kernel prior to restoring the original instruction, i.e., "*mov pc, lr*", that was substituted by the SMC instruction. Lastly, the processor exits the secure world and resumes execution starting from the restored instruction.

There are several advantages of the SPROBES mechanism. First, it is independent on the software running in the normal world. All the features used by SPROBES are natively provided by the ARM hardware, such as the SMC instruction and cross-world memory access. Second, an unforgeable state of the normal world can also be extracted from its hardware registers directly when an SPROBE is hit. Third, SPROBES are transparent to the normal world and thus do not require modifications to existing software. The original control flow in the normal world remains unchanged, so the operating system running in the normal world will not notice any overt differences caused by SPROBES. Fourth, SPROBES can break on any instruction in the normal world without any restrictions. This is because SPROBE does not cause any side effects on the normal world. Fifth, other than the SMC instruction, SPROBES are implemented in the secure world, so all of its code and data are isolated from the normal world. This limits the ability of a rootkit to affect any SPROBE execution. All these features combine to make SPROBES a powerful instruction-level instrumentation mechanism on TrustZone architecture.

## V. PROTECTING INTEGRITY OF KERNEL CODE

In this section, we propose an SPROBE placement strategy that can block all the approaches by which an adversary could violate the integrity of kernel code. With this placement strategy, we make a strong security guarantee that over the system's lifetime, any kernel rootkit cannot inject any code or modify approved kernel code. That is, even a rootkit in the normal world kernel is limited to running approved kernel code only. Further, since the SPROBES are inserted into kernel code, this placement strategy is sufficient to protect the SPROBES from modification as well.

**Attacker Model.** According to the problem stated in Section II, we base our work on the following attacker model. We assume there is at least one exploitable vulnerability in the kernel by which an adversary could hijack the control flow of the kernel, enabling that adversary to choose the address from which to run kernel code, such as return-oriented programming (ROP) attacks [22]. Defenses have been proposed that aim to reduce the adversaries' ability to leverage such attacks, such as address space layout randomization [19], [20]. In addition, researchers have proposed methods to counter ROP attacks, such as modifying the compiler to build a kernel without return instructions [24] and using VMM to enforce some degree of kernel-level control flow integrity [25], but these proposal still have some important limitations. Thus, we aim to limit adversaries with an installed rootkit to run only approved kernel code even though they may be capable of launching ROP-style attacks.

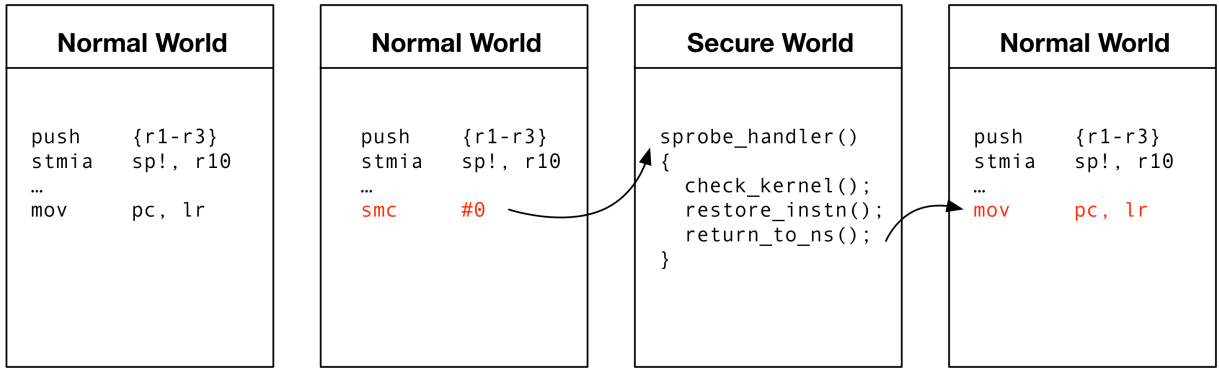

Fig. 3: World switches during SPROBE hit

**Trust Model.** In this paper, we make the following assumptions. In addition to trusting all the secure world code, we make the assumption that the normal world kernel code is free of rootkits prior to the execution of the first user-space process. That is, we assume that all rootkit threats originate from compromised root processes or any user-space process that has access to a kernel interface. We do not defend against malicious code running in the kernel prior to the first user-space process being initiated. In addition, we assume the use of hardware-based IOMMUs [26] to block malicious DMAs. Lastly, we trust the load-time integrity of kernel image is protected by utilizing technologies such as secure boot [16], [27].

**Placement Strategy Overview.** At a high-level, our strategy is for the secure world to configure the normal world kernel such that the SPROBES can mediate runtime access to memory management. This strategy combines enforcement of high-level management settings for ARM (e.g., keeping W⊕X enabled to enforce $S2$ and keeping the MMU enabled to enforce $S5$) with VMM-like breakpoints to control kernel memory access (e.g., verifying page table integrity to enforce $S3$ and validating page table updates to enforce $S4$), leveraging an ARM hardware feature to prevent the execution of user-space code from supervisor mode (e.g., to enforce $S1$). This strategy is implemented in three phases. First, the secure world must configure the kernel for booting in such a way that invariants $S1 - S5$ are satisfied initially (see Pre-boot Configuration below). Second, the kernel must ensure configuration of certain memory protections *prior to running the first user-space process* (see Boot Configuration below). Third, we describe how to place SPROBES such that they can implement policies necessary to enforce the invariants $S1 - S5$ throughout the kernel runtime (see Enforcing S1-S5 below).

**Pre-boot Configuration.** We require that the secure world has some specific knowledge about the normal world kernel memory in order to establish the enforcement of S1-S5 prior to booting. First, the secure world must know the base address that will be used for the kernel page table necessary to enforce $S3$. Second, the secure world must know the kernel code pages (see Approved Kernel Code below) in this page table, so it can validate the page table mappings (see Enforcing S3 below for the method), such as checking the correct page permissions necessary to enforce $S4$. Third, the secure world must write-protect the kernel page table to continue enforcement of $S4$. Other invariants will be established after booting the kernel.

**Approved Kernel Code.** In addition, the secure world must know that all of code page memory regions for the normal world kernel are approved for execution prior to booting. One issue with this assumption is that many kernels support loadable kernel modules (LKM), which may change the code in the kernel legitimately. However, for smartphones, LKM is not necessary because all the peripherals are known in advance and their drivers can be compiled along with the kernel. The Android 4.3 kernel on Nexus 4 is an example of a kernel where LKM is not supported. An alternative approach is to intercept *init_module* system call by inserting an SPROBE and record the memory location where module instructions are loaded. In addition, kernel modules may include security-sensitive instructions that may require SPROBE mediation as well. Thus, we assume that LKM is not supported in the normal world kernel and leave the more general problem for future work.

**Boot Configuration.** Once the normal world kernel is booted then further work is necessary to establish invariants $S2$ and $S5$. We require that W⊕X is set ($S2$) and the MMU is enabled ($S5$) prior to running the first user-space process. These are typically set early in the boot sequence, and SPROBES are placed to mediate access to these values as described below. If they are not set prior to changing the page table base value for the first time, indicating that a new process is running, then the secure world can halt the system. We note that although preventing the execution of user-space code in supervisor mode ($S1$) is not relevant until the first new process is initiated, we require the operating system to set the *Privileged eXecute-*

*Never* (PXN) bit[2] on all user-space pages before actually context switching to the first user-space process.

**Security Guarantees.** Using the above SPROBE placement strategy, we expect to obtain the information necessary to enforce invariants *S*3 and *S*4 over the kernel prior to boot, configure enforcement of *S*2 and *S*5 prior to running the first user-space process, and enforce *S*1 (prevent the kernel from running unprivileged code), *S*3 (enforce correct page table base addresses for the kernel and all processes), and *S*4 (prevent unauthorized modifications of the page table entries) over the kernel and user-space processes at runtime. We describe our approach to placing SPROBES to achieve these guarantees in the rest of this section.

**Enforcing S2.** As the WXN is a bit in the *System Control Register* (SCTLR), the rootkit would have to write to this register in order to turn off W⊕X protection. Recall that we assume that the WXN bit is set at initialization time. Therefore, the idea is if we could insert an SPROBE at every kernel instruction that writes to the SCTLR, we can trigger the secure world whenever a rootkit may attempt to turn off the protection. When such an SPROBE is invoked, it simply must block values that attempt to unset WXN to achieve *S*2. Since ARM has fixed length instructions, finding such instructions in the kernel binary is straightforward, particularly relative to x86.

One issue is that there are multiple control bits in the SCTLR, such as *Alignment Check Enable* and *MMU Enable*. This causes false sharing as updating a non-WXN bit in the SCTLR will also hit the SPROBE and trigger the secure world thus bring unnecessary overheads. We evaluate the performance impact of false sharing in Section VII.

**Enforcing S5.** Our solution does not insert additional SPROBES to prevent adversary from disabling the MMU. That is because the MMU Enable bit is in the same register, i.e., the SCTLR, as the WXN bit. All the SPROBES used to protect the WXN bit can also protect the MMU Enable bit. Therefore, *S*5 is satisfied. Given the fact that most operating systems do not disable MMU after it is turned on, it would be easy for the secure world to detect the presence of a rootkit if she tries to override the MMU Enable bit.

**Enforcing S3.** On ARM processors, the base address of page table is stored in a special register called *Translation Table Base Register* (TTBR). Similarly to above, by inserting an SPROBE at each instruction that writes to the TTBR, the secure world can fully mediate the operations that switch the page table, providing complete mediation of writes to this register.

Normally, to create separate address spaces for each process, the operating system assigns a different page table to each process. When a process is scheduled on processor, the operating system updates the TTBR with that process's page table base address. Thus, the secure world needs to be capable of ensuring that only valid page table bases are applied for each context switch. In order to do this, the secure world will have to maintain the integrity of the page tables. When a TTBR is asserted for the first time, the secure world must validate this new page table, ensuring that the addresses used are valid and checking compliance of the permission bits. For example, in Linux the kernel portion of the process's page table (i.e., addresses above `0xc0000000`) must be the same for each process page table and the approved kernel code pages must not be writable. In addition, double mapping (i.e., two virtual pages are mapped to the same physical frame) must not exist in the page table [28], particularly between a code page and a data page, otherwise the attacker can modify kernel code by writing to that data page. By restricting updates of TTBR to only valid page tables, we enforce invariant *S*3. Note that we only need to validate a page table the first time that we see its TTBR value because of the way we control page table updates to enforce *S*4 below. Note further that unlike protecting the WXN bit, the SPROBES to protect the TTBR will be hit in a regular manner due to process context switch. As a result, some performance overhead is fundamental to this enforcement.

**Enforcing S4.** Placing SPROBES to prevent the page table from being modified illegally by rootkits is the most challenging part in our solution. Unlike the cases above, which focus on mediating access to a special register, page tables are just normal memory, so consequently there are many usable instructions that can write to them. Inserting SPROBES at all of those instructions (e.g., *store*) is not practical because of performance overhead. But why do we want to instrument all memory store instructions if what we really want is to monitor updates to page tables, which are only a small portion of the whole address space? Therefore, instead, we apply write-protection over page tables, so that any table updates will generate a page fault. Then, if we insert an SPROBE into the page fault handler, the secure world will be triggered upon every table update. This is essentially how shadow page tables are implemented in virtualization, but we need to implement this mechanism utilizing ARM hardware.

Unlike Intel x86, ARM requires operating system to place its exception vector table at a fixed location, either `0x00000000` or `0xffff0000`, determined by *Vectors* bit in the SCTLR. Normally operating systems use `0xffff0000` as the exception base address because `0x00000000` is interpreted as NULL pointers. Because of this, the rootkit will not be able to re-define another exception vector table to bypass the SPROBE in the page fault handler. It is also worth noting that exception vectors are code rather than data, so they cannot be written by a rootkit without changing the code page permissions, which is what this SPROBE enforces. Each vector contains an instruction (normally a branch instruction), and the instruction is executed once a corresponding exception happens (e.g., page fault). To illustrate, we show an example of exception vector table in assembly code as Listing 1. For this listing, we propose to insert an SPROBE at the start of the function `abort_handler`, which services page faults. Therefore, given that the adversary can neither re-define a new exception vector table, nor modify the current one, updates to the normal world's page tables are fully mediated by the secure world.

When the secure world is triggered due to a page fault in the normal world, there are three situations worth discussing, and the flow chart of SPROBE handling is shown as Fig. 4. First, if the page fault is intended such as due to copy-on-write fork, we simply return to the exception handler to let

---

[2]The PXN bit is a permission bit in page table entries that determines whether a memory region is executable from privileged modes

Listing 1: Exception Vector Table

```
; we are at 0xffff0000 now
exception_vector_table:
reset:
    b      init             ; system boots
undefined:
    b      undefined
supervisor:
    b      syscall_start    ; system call
prefetch_abort:
    b      abort_handler    ; page faults
data_abort:
    b      abort_handler    ; page faults
unused:
    b      unused
irq:
    b      irq_handler      ; interrupts
fiq:
    b      fiq              ; not used
```

the kernel deal with this exception. Second, if the page fault is caused by a legitimate table update like mmap system call, the secure world would first emulate the behavior of table update and then return to the faulted instruction as if nothing happens. By returning to the instruction that causes the page fault instead of the next instruction in the exception handler, we make the whole procedure invisible to the operating system since it does not expect such an exception. Third, if the page fault is caused by either (1) making a page writable that maps to a physical frame containing kernel code, or (2) making a page executable that maps to any other physical frame, the secure world should block such page table modifications and trigger rootkit detection mechanisms. Therefore, $S4$ is satisfied.

**Enforcing S1.** Since all page table entries are protected from modification by enforcing $S4$, the PXN bits set prior to the installation of rootkit are safe from modification and the secure world can prevent the rootkit from creating new mappings in which the PXN bits are unset. In addition, $S3$ can reject a page table whose PXN bits are not properly set for new page tables, even those created by a compromised kernel. Thus, all the page tables must have the PXN bits set for all user-space pages, satisfying $S1$.

We enforce all five of these invariants by mediating a fixed set of instruction types. An implicit assumption behind our design is that the rootkit cannot jump into the middle of an instruction to discover unintended sequences. Fortunately, ARM, like most other RISC machines, does not support misaligned instruction fetch, defeating this possibility completely.

In summary, by enforcing invariants from $S1$ to $S5$, we mediate all the ways that a party, either the legitimate kernel or a malicious adversary, controls the system's virtual memory environment. $S5$ ensures the MMU is always on, which serves as a foundation for the rest of protections. $S3$ further limits the usable page tables to a small set, preventing the adversary from setting the page table base to an arbitrary value. $S4$ protects the integrity of page table entries, thwarting any attempts to modify kernel code and thus protecting the SPROBES from unauthorized removal. As a supplement, $S1$ and $S2$ disallow execution over injected instructions and assure the enforced mediation is not bypassable. Overall, the five invariants restrict the adversaries to approved code pages, indirectly protecting any SPROBES inserted into these pages for enforcement purposes.

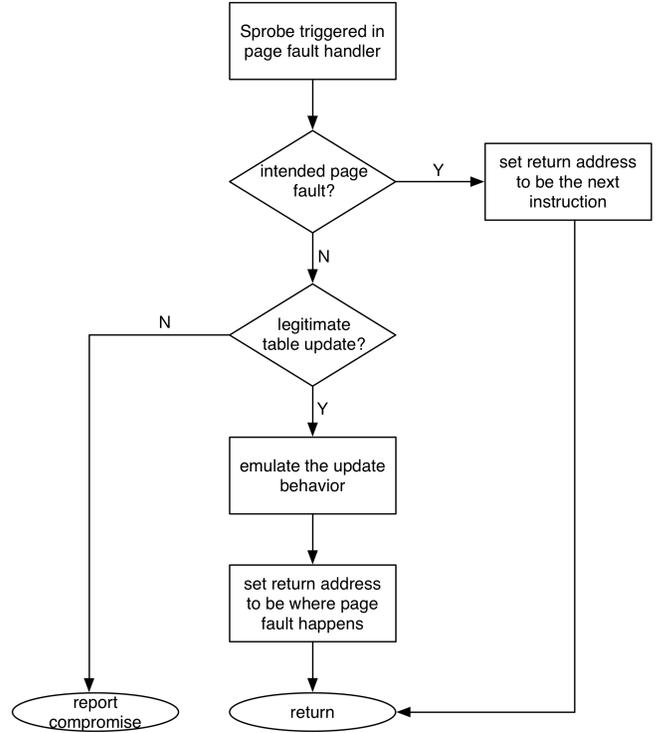

Fig. 4: SPROBE handling on data abort

VI. IMPLEMENTATION

We implement a prototype of SPROBES on top of a Cortex-A15 processor emulated by Fast Models 8.1 emulator [29] from ARM. Although SPROBES do not rely on the software (e.g., kernel) running in the normal world, to have a proof of concept of our solution, we run Linux 2.6.38 in the normal world as a case study. We build Linux from source code using the GNU ARM bare metal toolchain (e.g., arm-none-eabi). This toolchain is mainly for building applications for the ARM architecture without any operating system support. To make the SPROBES implementation simpler and more efficient, we extract all the necessary kernel information before hand. For example, we use *objdump* to inspect the address space layout of kernel code to identify the locations for SPROBES.

One key step in SPROBES implementation is to substitute the target normal world instructions with SMC instructions. In most cases, the MMU is enabled in the normal world, so the secure world can only see its virtual addresses. To access a given virtual address in the normal world, the secure world needs first to translate it to the physical address and then create a corresponding mapping in its own page table. Note that the physical address space of the two worlds can overlap, so the page table entry has an NS bit to indicate which world the physical address is from. To convert a virtual address to a physical address, the Cortex-A15 architecture has an external coprocessor (CP15) that can complete such a translation, which avoids manually walking the page table in the normal world.

In order to enforce W⊕X protection, we disassemble the text section of kernel image and identify all instructions that write to the SCTLR. Then, in the secure world, we hardcode the addresses of those instructions and insert SPROBES after the Linux kernel is loaded in the normal world.

Similarly, to protect the TTBR, we scan the disassembled text section of kernel image and record all instructions that can write to the TTBR. However, ARM processors support more than one page table active (two in maximum) at a time, and one is determined by the TTBR0 while the other is determined TTBR1. The TTBRs are used together to determine addressing for the full address space. Which table is used for what address range is controlled via the *Translation Table Base Control Register* (TTBCR). In the actual implementation of Linux, by setting the TTBCR to a fixed value, it uses only one page table throughout its lifetime. However, to prevent adversary from enabling the second page table, we still need to insert SPROBES at those instructions that write to the TTBCR.

Protecting the page table requires the secure world to modify page permissions of the normal world. As stated in Section V, the secure world sets the page table memory to be read-only, so that any page table updates cause page faults. The implementation is actually similar to the mechanism to synchronize guest page tables and *shadow page tables* that are used by the VMM [30], [31]. Since we also insert an SPROBE to the page fault handler, the secure world would be triggered on every page fault. The chain connecting the page table updates to the secure world does not give the normal world software any opportunity to interfere with this function, as none of the normal world instructions can be executed in the middle. This requires us to insert an SPROBE at the very first instruction of the exception handler. There are two benefits by doing so: (1) it ensures no normal world instruction is executed before the secure world gains the control and (2) in those cases where the operating system does not expect such a page fault (e.g., mmap system call), it minimizes the effects on the processor state of the normal world and thus makes state recovery (i.e., restore the register values before the page fault) easier.

## VII. EVALUATION

### A. Security Analysis

In this section, we evaluate the security of our solution by summarizing how our design has achieved the goal of protecting kernel code integrity for Linux 2.6.38.

For illustration, we list the 12 SPROBES necessary to implement the placement strategy for Linux 2.6.38 in four groups:

- **Type #1**: The 6 SPROBES that protect the SCTLR containing the WXN and MMU Enable bit.

- **Type #2**: The 4 SPROBES that protect the TTBR containing the base address of the page table.

- **Type #3**: An SPROBE that protects the TTBCR to enforce usage of only one TTBR (i.e., TTBR0), as required by Linux.

- **Type #4**: The SPROBE that is inserted at the first instruction of the page fault handler.

At a high level, we demonstrate our design can effectively reduce the adversary to approved kernel code following these steps. We first show that the adversary has to change the memory environment, in order to execute injected or modified code in kernel space. Then, we illustrate how the combination of listed four types of SPROBES enables the secure world to detect all changes to the normal world memory environment. Finally, we claim that no memory environment changes for a malicious purpose can bypass the checks in the secure world.

First, according to the Boot Configuration in Section V, the only possible attack against the integrity of kernel code without tampering with the virtual memory environment is to modify the kernel image file, so that the system would be in a "compromised" state the next time it is loaded into memory. We foil such attacks by utilizing technologies like secure boot [16], [27] as assumed in the Trust Model in Section V. Note that checking the integrity of kernel image file is sufficient to ensure load-time kernel integrity as kernel loading is at the very beginning of a boot sequence, before which no adversary is assumed to have access to the system.

Second, we claim that modifying the virtual memory environment will always be captured by the secure world regardless of its purpose. In essence, a virtual memory environment is uniquely defined by the active page tables as long as the MMU is on[3]. So, if not considering the cases where the MMU is disabled, modifying the virtual memory environment is just equivalent to modifying the active page tables. To accomplish this, the attacker may either (a) switch to a set of page tables that are under her control or (b) modify page table entries in place. However, (a) would cause the hit of Type #2 and/or Type #3 SPROBES while (b) triggers Type #4 SPROBES since page tables are write-protected. Alternatively, the attacker can simply disable the MMU, accessing physical memory with no restrictions. Similarly, such operation will be trapped to the secure world as well because of Type #1 SPROBES.

Finally, we show that altering the virtual memory environment for a malicious purpose will not bypass the checks in the secure world. To achieve this, we need to draw a clear boundary between legitimate and malicious operations on the memory settings. To begin with, triggering Type #1 SPROBES, either by disabling the MMU or the W⊕X protection, is a clear indication of system compromise in the normal world because generally an operating system (e.g., Linux) will not turn any of them off after they are on. When modifying the active page tables either by (a) or (b), triggering Type #2/#3 or Type #4 SPROBES, we enforce the same permission settings as shown in Fig. 5: consisting of non-executable kernel data, unwritable kernel code, and non-executable user pages. Any attempt to violate this configuration will be regarded as a malicious operation. Further, in order to ensure that *physical kernel code frames* are protected by managing the permission settings on virtual pages, we ensure that there is a *fixed one-to-one* mapping between kernel code pages and their corresponding physical frames. In addition, by forbidding double mappings, we rule out the possibilities of modifying a physical code frame through a virtual data page, the combination of which ensures only a set of unmodified physical frames have been executed since system starts, which is our goal in this paper.

---
[3]Though the permission settings can be enhanced through bits like SCTLR.WXN.

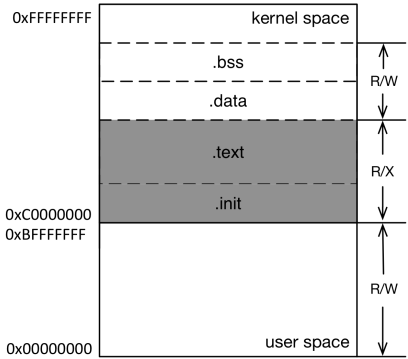

Fig. 5: Normal world virtual memory layout

*B. Performance Evaluation*

In this section, we evaluate the performance impact of SPROBES on the normal world execution. The limitation of this performance evaluation our proof-of-concept was implemented on the Fast Models emulator, but Fast Models does not model accurate cycle counts, instead tracking instruction counts. For example, arithmetic operations (e.g., add) and memory accesses (e.g., load) all take the same amount of time, one instruction. This makes running benchmarks on Fast Models meaningless, as the perceived overheads of software running on the model differs from real-world software.

To obtain a general understanding of how much overhead SPROBES incur, we count the number of instructions instead. An SPROBE hit causes 5611 more instructions (including the original SMC instruction) to be executed in the secure world. Note that the number of instructions is a very coarse-grained measurement as it does not take microarchitectural events into account.

We run Linux 2.6.38 in the normal world with 28 startup processes including 4 daemon processes, 1 interactive process and 23 kernel threads. We run a shell script that invokes the write system call in a loop as the workload. We measure the individual cost for SPROBES of the four different types. To understand how frequently those SPROBES are hit in Linux 2.6.38, we use hit frequency, the average number of elapsed instructions between two consecutive hits, as the metric. We list our results in Table I. Both Type #1 and Type #3 SPROBES are not hit after the booting, which means enforcing $S2$, $S3$ and $S5$ incurs negligible runtime overheads. Type #2 SPROBES are hit on each context switch. On average 313,836 instructions are executed between each Type #2 SPROBE hit, contributing to 2% of the instructions executed. We further measure the Type #4 SPROBES during boot stage when page updates are the most intensive. The result turns out Type #4 SPROBES are hit in every 22,424 instructions. After the kernel is setup, the hit frequency goes down to every 85,982 instructions.

Since smartphone boot-times can be performance-critical, SPROBES overhead may be an issue. Perhaps restricting the code executed at boot-time to only trusted code is necessary to achieve performance goals. However, this may introduce a window of attack for adversaries. Normal runtime overheads incurred by SPROBES may be close to acceptable for the protection offered, given that only 2 types of SPROBES are hit and incur fewer than 10% of the instructions executed. We note that current VM introspection methods trap all page faults as well, so the number of traps would be the same, although the overhead for TrustZone may be higher.

| SPROBE Type | #1 | #2 | #3 | #4 |
|---|---|---|---|---|
| Hit Frequency | N/A | 313,836 | N/A | 85,982 |

TABLE I: Hit frequency of different types of SPROBES

Revisiting the false sharing issue stated in Section V, the fact that Type #1 SPROBES are never hit after initialization demonstrates that false sharing can be eliminated to an acceptable range in a real implementation.

## VIII. RELATED WORK

The main focus of this paper is to enforce kernel code integrity. SecVisor [10] and NICKLE [11] are two VMM-based approaches proposed to protect lifetime kernel code integrity. SecVisor is a tiny hypervisor (e.g., fewer than 2,000 SLoC) that restricts the code running in kernel space. It is achieved by virtualizing physical memory to allow SecVisor to exclusively set hardware memory protection over kernel memory, which is independent on the memory protection in the guest machine. In addition, SecVisor also checks certain kernel states upon mode switch, e.g., return from a system call. NICKLE is implemented as part of a virtual machine monitor which allows only authenticated code to be run. Besides the standard memory for running the operating system, it also maintains a separate physical memory region called *shadow memory* and stores a duplication of authenticated code in this area. The VMM enforces the guest kernel cannot access the shadow memory so the integrity of code within this region is protected. During runtime, NICKLE transparently routes instruction fetches to this shadow memory and thus blocks any attempt to execute unauthenticated instructions.

Petroni *et al.* proposed a state-based control flow integrity (SBCFI) monitor [32] that detects kernel rootkits by periodically snapshotting the memory in a VM. It relies on the assumption that possible desired states of memory are enumerable. Because of this, in practice, the detection is mainly effective on invariant (e.g., kernel code) or enumerable contents (e.g., some global function pointers). However, a sophisticated attack can remain undetected if it only occurs between two snapshots, which is a limitation of asynchronous detection.

Garfinkel *et al.* first proposed a VMM-based monitor, called Livewire [2], to mainly protect system invariants. The Livewire system introduced the VM introspection technique and is able to check the integrity of kernel code and verify function pointers with fixed values (e.g., system call table).

Zhang *et al.* proposed a coprocessor-based kernel monitor [12]. Though it may not have full access to the state of host processor (e.g., registers), by using an additional piece of hardware like PCI cryptographic coprocessor, the solution improves the overall system performance comparing with VMM-based monitoring. Comparably, Petroni *et al.* further proposed a prototype of coprocessor-based kernel monitor called Copilot [13], which periodically checksums invariant memory regions and sends reports to a remote administration station. The autonomous subsystem provided by the coprocessor is similar to TrustZone, although the secure world is much more powerful as detailed in Section III.

Wang *et al.* built a hardware-assisted tampering detection framework called HyperCheck for VMMs [14]. HyperCheck utilized Intel's System Management Mode (SMM) to reliably check the state of the VMM and securely communicate to a remote administration server.

## IX. CONCLUSION

In this paper, we have presented the design and implementation of SPROBES, an instrumentation mechanism that enables the secure world to introspect the operating system running in the normal world. To protect the SPROBES, we identify a set of five invariants and present an informal proof of how 12 SPROBES enforce these invariants comprehensively. By enforcing all five of these invariants, we make a security guarantee that only approved kernel code is executed even if the kernel is fully compromised.

## X. ACKNOWLEDGEMENT

This material is based upon work supported by the National Science Foundation under Grant No. CNS-1117692. Research was sponsored by the Army Research Laboratory and was accomplished under Cooperative Agreement Number W911NF-13-2-0045 (ARL Cyber Security CRA). The views and conclusions contained in this document are those of the authors and should not be interpreted as representing the official policies, either expressed or implied, of the Army Research Laboratory or the U.S. Government. The U.S. Government is authorized to reproduce and distribute reprints for Government purposes notwithstanding any copyright notation here on. Finally, we are grateful for many technical discussions with Jason Chiang and Rick Porter and their continuous help during this project.